\documentclass[acmsmall]{./acmart}

\usepackage{colortbl}
\usepackage{soul}
\usepackage{enumitem}
\usepackage{fontawesome5}
\usepackage{subcaption}

\newcommand{\pc}[1]{(#1 participants)}

\definecolor{amber}{rgb}{1.0, 0.75, 0.0}
\newcommand{\hlc}[1]{\sethlcolor{lightgray!80}\hl{\small \texttt{#1}}}

\AtBeginDocument{%
  \providecommand\BibTeX{{%
    \normalfont B\kern-0.5em{\scshape i\kern-0.25em b}\kern-0.8em\TeX}}}

\newcommand{\qp}[2]{{\small \textsl{``#1'' (#2)}}}
\newcommand{\qpt}[2]{{\footnotesize \textsl{``#1'' (#2)}}}
\newcommand{\qpwi}[1]{{\small \textsl{``#1''}}}
\newcommand{\pidf}[1]{{\small $#1$}}
\newcommand{\pidfv}[1]{{\small $#1_{v}$}}

\newcommand{\iconformatting}[1]{(\hspace{0.05em}\raisebox{-0.1em}{#1}\hspace{0.05em})}
\newcommand{\uiconformatting}[1]{\hspace{0.05em}{\footnotesize{#1}}}
\newcommand{\capiconformatting}[1]{({\small{#1}})}
\newcommand{\content}{\iconformatting{\faClipboardCheck}}
\newcommand{\presentation}{\iconformatting{\faSpellCheck}}
\newcommand{\userexp}{\iconformatting{{\faSmile[regular]}}}

\newcommand{\utw}{\uiconformatting{\faUserEdit}}
\newcommand{\udev}{\uiconformatting{\faUserCog}}
\newcommand{\upm}{\uiconformatting{\faUserTie}}
\newcommand{\uqa}{\uiconformatting{\faUserCheck}}
\newcommand{\uext}{\uiconformatting{\faUsers}}

\newcolumntype{L}[1]{>{\raggedright\arraybackslash}p{#1}}

\acmConference[GROUP '27]{ACM International Conference on Supporting Group Work}{January 10--13, 2027}{St. Simons Island, GA, USA}
\setcopyright{none}
\makeatletter
\patchcmd{\@mkbibcitation}{In \textit{\@acmBooktitle}}{To appear in \textit{\@acmBooktitle}}{}{
  \typeout{Patch of \string\@mkbibcitation\space failed}
}
\patchcmd{\ps@standardpagestyle}
  {\fancyfoot[RO,LE]{\footnotesize \@journalName, Vol. \@acmVolume, No.
     \@acmNumber, Article \@acmArticle.  Publication date: \@acmPubDate.}}
  {\fancyfoot[RO,LE]{}}
  {}{\typeout{Patch of \string\ps@standardpagestyle\space failed}
  }

\renewcommand\@formatdoi[1]{}
\makeatother
\renewcommand\footnotetextauthorsaddresses[1]{}
\renewcommand\footnotetextcopyrightpermission[1]{}

\begin{document}
\title[A Second Set of Eyes]{``A Second Set of Eyes'': The Process and Challenges of Software Documentation Review} 

\author{Avinash Bhat}
\authornote{Corresponding author: avinash.bhat@mail.mcgill.ca}
\email{avinash.bhat@mail.mcgill.ca}
\affiliation{%
  \institution{McGill University}
  \city{Montreal}
  \country{Canada}
}

\author{Ian Arawjo}
\email{ian.arawjo@umontreal.ca}
\affiliation{%
  \institution{Université de Montréal}
  \city{Montreal}
  \state{Quebec}
  \country{Canada}
}

\author{Disha Shrivastava}
\email{shrivasd@google.com}
\affiliation{%
  \institution{Google DeepMind}
  \city{London}
  \country{United Kingdom}
}

\author{Jin L.C. Guo}
\email{jguo@cs.mcgill.ca}
\affiliation{%
  \institution{McGill University}
  \city{Montreal}
  \state{Quebec}
  \country{Canada}
}

\begin{teaserfigure}
    \centering
    \includegraphics[width=\linewidth]{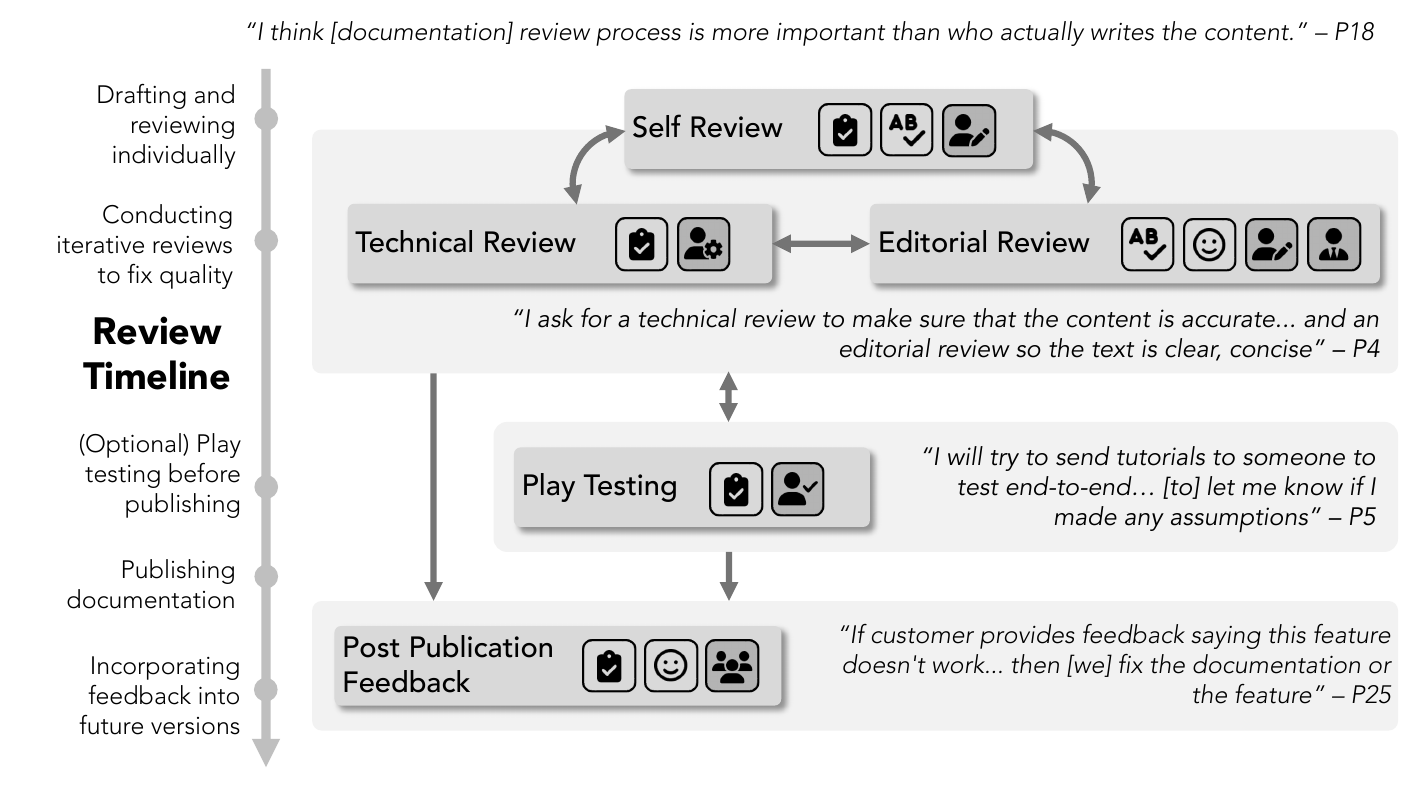}
    \caption{An overview of the documentation review process. In each stage, we indicate the quality categories reviewed: Content \capiconformatting{\faClipboardCheck}, Presentation \capiconformatting{\faSpellCheck}, and User Experience \capiconformatting{\faSmile[regular]} and the practitioners involved: Technical Writers \capiconformatting{\faUserEdit}, Developers \capiconformatting{\faUserCog}, Program Managers \capiconformatting{\faUserTie}, QA \capiconformatting{\faUserCheck}, Users \capiconformatting{\faUsers}.} 
    \label{fig:process_overview}
    \Description{}
\end{teaserfigure}

\begin{abstract}
Organizations assign documentation work to technical writers, yet the knowledge required to produce it is distributed across developers, managers, and other practitioners. Prior work has established quality criteria for judging "good" documentation, but it has not examined how practitioners bring that expertise to improve documentation quality or the challenges they face in doing so. Through semi-structured interviews with experienced technical writers ($n=31$) from different organizations, our work reveals the individual and collaborative effort required to maintain documentation quality.  We identify five distinct stages of the documentation review process: \textit{self review}, \textit{technical review}, \textit{editorial review}, \textit{play testing}, and \textit{post-publication feedback}. Each stage draws on practitioners with distinct expertise to address quality across content, presentation, and user experience. Our findings surface organizational and technical challenges writers face in recruiting expert reviewers, navigating development timelines, and contending with tools not specifically designed for documentation workflows. Our work positions documentation review as a crucial yet understudied site of collaborative work and opens new research and design directions for process improvement and tool support.
\end{abstract}

\maketitle
\section{Introduction}
Software documentation supports diverse consumers of software technology, including developers, end users, and maintainers, providing instructions of the software usage and explanations on the technical details of its implementation~\cite{forward_relevance_2002, lethbridge_how_2003}. High-quality documentation improves software adoption~\cite{dagenais_creating_2010, sohan_study_2017}, reduces maintenance costs~\cite{arisholm_impact_2006, plosch_value_2014, kazman_evaluating_2016}, facilitates team member on-boarding and prevents knowledge loss when people leave~\cite{robillard_turnover_2021}. Conversely, poor documentation creates both immediate and long-term impacts on software project success, such as user abandonment~\cite{uddin_how_2015, robillard_field_2011}, delayed product release~\cite{garousi_usage_2015, maalej_patterns_2013}, and increased support costs~\cite{aghajani_software_2019, garousi_evaluating_2013}.

Due to its significance, prior work has investigated what constitutes high-quality documentation and identified quality attributes, such as accuracy, completeness, and usability, and methods for evaluating them~\cite{zhi_cost_2015, treude_beyond_2020, aghajani_software_2019, tang_evaluating_2023}. For instance, \citet{treude_beyond_2020} organized quality dimensions into structure, content, and style. Various quality frameworks have been proposed, using metrics to quantify documentation staleness relative to code changes~\cite{plosch_value_2014, mcburney_automatic_2015}, models to identify documentation smells~\cite{khan_automatic_2021}, and tools to measure structural issues~\cite{tang_evaluating_2023}.

Despite these works on understanding documentation quality, studies consistently reveal problems associated with it, including insufficient content and obsolete information~\cite{uddin_how_2015, aghajani_software_2019, aghajani_software_2020}, even when having dedicated documentation resources~\cite{maalej_patterns_2013, garousi_evaluating_2013}. This disconnect suggests a fundamental issue: \textbf{we know \emph{what} high-quality documentation looks like, but not \emph{how} it is achieved in practice}. Existing work focuses predominantly on defining quality attributes rather than understanding the \textit{processes} used to achieve these standards. \citet[p. 195]{zhi_cost_2015} explicitly identified this limitation, stating ``stronger empirical evidences are still needed to enhance the understanding and to establish profound theories'' regarding documentation workflows. 

The recent trend of automatically generating software documentation~\cite{mcburney_automatic_2014, mcburney_automatic_2015}, especially through large language models (LLMs)~\cite{khan_automatic_2022, luo_repoagent_2024, dearstyne_supporting_2024, bhat_do_2024}, makes this gap increasingly urgent. Practitioners struggle with new quality challenges like hallucinations and consistency issues introduced by these systems~\cite{huang_survey_2025, ji_survey_2023, alshahwan_assured_2024}. However, without understanding how quality is actually achieved, we struggle to provide adequate support for documentation processes in an era of rapidly evolving tools.

To address this gap, we investigated documentation quality practices through semi-structured interviews with 31 experienced technical writers from different organizations, ranging from startups to enterprise companies and open-source projects, spanning industries including data platforms, security, hardware, military, and aviation. We focused on technical writers because existing research that touches on documentation has focused primarily on developers' perspectives~\cite{tang_evaluating_2023, aghajani_software_2020}, overlooking the systematic approaches employed by technical writers who possess specialized expertise in approaching and assessing documentation quality~\cite{treude_beyond_2020}. In particular, we investigated two research questions: \newline
\noindent \textbf{RQ1: How is documentation quality achieved in practice? (\S\ref{sec:doc_quality})} Our findings reveal that practitioners achieve documentation quality through a \textbf{collaborative review process}. While demonstrating variance in different organizations, this process generally includes five stages (see Fig.~\ref{fig:process_overview}); each involves practitioners with specialized expertise who focus on specific quality goals across three categories: \emph{Content} \content, \emph{Presentation} \presentation, and \emph{User Experience} \userexp. The review process is also instrumented with extensive tooling to prevent issues during writing, support collaborative review stages, and collect feedback after publication. 

\noindent \textbf{RQ2: What are the challenges to achieving high quality documentation through reviews? (\S\ref{sec:challenges_review})} We identify two categories of challenges. Process challenges include poorly planned documentation cycles, competing stakeholder priorities, and expertise gaps between writers and reviewers. Tooling challenges include missing documentation-specific tools, the maintenance burden documentation tooling imposes, and LLMs' unreliability and security risks. These findings reveal the gaps between how documentation review actually operates and current organizational support.

Our work makes the following contributions: (1) We characterize the documentation review process and the tools involved, establishing how practitioners coordinate documentation quality through multiple experts, (2) We identify the specific challenges technical writers face during review, revealing significant gaps in process and tools that explain why quality issues persist and how they develop workarounds, and (3) Informed by these findings, we derive a design space that characterizes efforts toward quality through the intervention timing and responsibility of decision making dimensions, which enables both researchers and industry professionals to systematically explore alternatives to current approaches and identify improvement opportunities. These contributions provide an empirical understanding of the documentation quality process and insights that explain where existing systems might fail and how to better support quality achievement through organizational effort and technological advancement.

\section{Related Work}
Achieving documentation quality requires practitioners from different roles to evaluate the same artifact, and tools to support that process. We review relevant HCI, CSCW and Software Engineering (SE) research: studies on collaboration which examine how practitioners with different expertise evaluate shared work and studies on reviewing which examine code review as a collaborative process and the tools built to support it.

\subsection{Collaboration in Software Engineering Teams}
\label{subsec:collab_in_se}
Researchers have studied collaboration in SE across several settings: distributed projects coordinating across geographic and cultural boundaries~\cite{herbsleb_splitting_1999, espinosa_team_2007}, colocated project teams~\cite{curtis_field_1988, begel_coordination_2009}, and cross-disciplinary collaborations involving practitioners such as data scientists, UX designers, and technical communicators~\cite{nahar_collaboration_2022, subramonyam_solving_2022, feng_when_2025, li_crossdisciplinary_2017}. Across these settings, research has documented how artifacts and processes cannot cover every coordination need that arises in practice~\cite{herbsleb_splitting_1999, begel_coordination_2009}, how expertise is distributed unevenly across practitioners~\cite{curtis_field_1988, faraj_coordinating_2000, espinosa_team_2007, kotlarsky_are_2015}, and how practitioners from different fields judge the same work differently, with no agreed standard for whose judgment takes precedence~\cite{nahar_collaboration_2022, subramonyam_solving_2022, feng_when_2025, li_crossdisciplinary_2017, whitehead_collaboration_2007}.

Filling the coordination gaps requires more than having the right expertise on a team; someone must actively identify where expertise resides and direct it accordingly~\cite{faraj_coordinating_2000, herbsleb_splitting_1999, begel_coordination_2009}. CSCW uses the term \emph{articulation work} to describe the effort of aligning the distributed contributions in a cooperative task so that they are coherent together~\cite{strauss_articulation_1988, corbin_articulation_1993}; articulation work can impose a significant overhead on whoever ends up performing it~\cite{schmidt_taking_1992}. Teams that do this outperform those who merely possess the expertise~\cite{faraj_coordinating_2000}. However, practitioners from different disciplines often cannot tell who across the team holds the expertise their work requires~\cite{kotlarsky_are_2015, espinosa_team_2007}. This coordination defaults to practitioners whose expertise bridges across teams~\cite{curtis_field_1988, damian_role_2013, malone_interdisciplinary_1994, crowston_coordination_1997, grinter_supporting_1996}, often producing informal artifacts such as sketches, annotated diagrams, and ad hoc documents to make their work understandable to collaborators from other fields~\cite{barrett_boundary_2010, slattery_undistributing_2007}. This work leaves no trace in any document or system record~\cite{star_layers_1999, kross_orienting_2021, piorkowski_how_2021}; neither do the conversations in which practitioners negotiate whose standards apply~\cite{passi_making_2020, feng_when_2025}. Because none of it produces anything visible, practitioners who do it report that it goes unrecognized and uncompensated~\cite{deng_investigating_2023, li_crossdisciplinary_2017}. Such \emph{invisible labor} accounts for roughly half the work in open source ecosystems~\cite{meluso_invisible_2025, star_layers_1999}.

We investigate collaboration in software documentation review context, which has not been studied previously. Unlike the settings examined in prior work, most participants in documentation review consider it as a secondary task which presents distinct challenges for technical writers who coordinate the process. We use articulation work to describe what these writers do, and show that the coordination they perform is invisible to the colleagues who benefit from it.

\subsection{Reviewing in Software Engineering}
\label{subsec:reviewing_in_se}
The reviewing process in SE has predominantly been used to refer to code review. Code review has evolved from formal inspections~\cite{fagan_design_1976} to modern lightweight practices~\cite{rigby_convergent_2013, eldh_code_2024} effective at preventing defects and security issues~\cite{bavota_four_2015, thompson_large-scale_2017}. This practice has attracted substantial research attention, covering practitioner expectations and outcomes~\cite{bacchelli_expectations_2013}, reviewer competencies~\cite{wurzel_competencies_2023}, accountability mechanisms~\cite{alami_accountability_2025}, and social dynamics such as interpersonal conflicts~\cite{goncalves_interpersonal_2022,goncalves_constructive_2024, qiu_detecting_2022} and pushback~\cite{egelman_predicting_2020}. Most of this work shows that review depends on the exchange between the author and the reviewer rather than on the artifact alone. Reviewers spend more effort understanding a change than finding faults in it~\cite{bacchelli_expectations_2013, ebert_confusion_2019}, and what lets them understand it is the rationale the author supplies~\cite{pascarella_information_2018}. The author's standing and organizational position also affect whether a patch is reviewed and how quickly~\cite{baysal_influence_2013}, and developers name reviewer availability as their central obstacle to review quality~\cite{kononenko_code_2016}. Documentation review depends on similar exchange between authors and reviewers, but under more inflexible conditions since reviewers belong to other roles such as developers, product managers, and marketing or legal, and therefore judge documentation by the standards of those roles. We study how the review proceeds under these conditions, which has not been examined.

Tools that support review in software engineering, again, have been developed primarily around code. In a systematic review of modern code review literature, \citet{davila_systematic_2021} found that tools exist for recommending reviewers, visualizing code changes, and supporting broader workflow interactions such as annotating code, prioritizing review requests, and tracking review history. More recent work has extended this further into automating the review activity itself, training models to recommend code changes and implement reviewer comments~\cite{tufano_towards_2021, tufano_code_2024}. Other work aims to improve reviewer experience, for example by automatically detecting toxic comments~\cite{sarkar_automated_2023}. Documentation review has no such tooling, and we describe the workarounds writers rely on instead.

\subsection{Documentation Quality and Practice}
\label{subsec:doc_quality_practice}
Software engineering literature points at multiple aspects of documentation quality, including accuracy, up-to-date content, completeness, and findability~\cite{treude_beyond_2020, tang_evaluating_2023, plosch_value_2014, zhi_cost_2015}. However, the tooling predominantly addresses accuracy. Most tools detect inconsistencies between code and documentation, treating source code as ground truth; DocRef~\cite{zhong_detecting_2013}, FreshDoc~\cite{lee_automatic_2021}, and Doc2OracLL~\cite{hossain_doc2oracll_2025} each target this concern. A smaller set of tools supports collecting feedback from community members after publication~\cite{watson_tool_2016, mysore_porta_2018}. However, no existing work addresses the remaining dimensions or the complete review process. Our work identifies these gaps and provides the empirical foundation that tool designers need, opening documentation review as a research area. 

In an organizational setting, practitioners report that projects lack good quality documentation~\cite{stettina_necessary_2011}. However, effort spent on documentation is usually low. In continuous development, productivity is mostly measured by the working software delivered, so time spent on documentation is given low priority; this results in the documentation drifting out of sync with the code~\cite{theunissen_mapping_2022}. Closest to our work, \citet{geiger_types_2018} reported that contributors to open-source libraries were demotivated by how little credit documentation work receives compared to code when creating and maintaining those projects. In our context, writing documentation is an assigned responsibility, so writers are accountable for its quality, while the reviewers they depend on to achieve that quality are under no obligation to help, and we describe how writers manage this.

\section{Research Design}
\label{sec:method}
Understanding how documentation quality is achieved in practice requires investigating the organizational processes through which practitioners establish and maintain quality standards. We therefore conducted semi-structured interviews with experienced technical writers and analyzed them using qualitative methods~\cite{seaman_qualitative_1999}. The study design was approved by the research ethics board of the authors' university.

\subsection{Participants and Recruitment}
\label{subsec:recruitment}
We recruited participants with experience in both writing and reviewing technical documentation by advertising the study in two popular public communities frequented by tenured technical writers: Write The Docs Slack~\cite{write_write_2025} and the Technical Writer Forum on LinkedIn~\cite{raghuram_technical_2025}. Our screening process filtered for familiarity with documentation review processes. In addition, to verify that participants had meaningful documentation experience, we asked each to share at least one publicly available example of their documentation work, which we manually checked for sufficient length, subject matter depth, and inclusion of instructional resources such as code snippets or GUI screenshots.

We recruited technical writers because they orchestrate the documentation review process. The coordination, such as locating reviewers or reconciling conflicting feedback from different reviewers, is often invisible; managers and developers see only the request that reaches them rather than the work of technical writers that produced the request. This is characteristic of \emph{articulation work}, which typically goes unrecorded and unnoticed by those who benefit from it~\cite{strauss_articulation_1988, star_layers_1999} as we discuss in \S\ref{subsec:invisible_work}. Writers are therefore the only practitioners positioned to report on this coordination, which is what makes their account the appropriate one for our research questions.

\begin{table}[tbp]
  \centering
  \rowcolors{2}{gray!15}{white}
  \small
  \caption{Demographics of the 31 participants. The parenthesis has the corresponding participant count. A detailed table is provided in the supplementary material.}
  \begin{tabular}{p{0.31\columnwidth}>{\raggedright\arraybackslash}p{0.59\columnwidth}}
  \toprule
  \textbf{Demographics} & \textbf{Breakdown} \\
  \hline
  \textbf{Recruitment Source} & LinkedIn (19), Write The Docs Slack (12) \\
  \textbf{Professional SE Experience} & 1-4 years (9), 5-10 years (11), 11-15 years (7), \mbox{>15 years (4)} \\
  \textbf{Current \mbox{Occupation}} & \textbf{Primary Reported Role}: Technical Writer (28), \mbox{Manager (2)}, \mbox{Software Engineer/Developer (1)}\\
  & \textbf{Additional Reported Roles}: Manager (1), \mbox{UX Writer (1),} \mbox{Support Engineer (1)} \\
  \textbf{Documentation \mbox{Experience}} & 1-4 years (5), 5-10 years (14), 11-15 years (7), \mbox{>15 years (5)} \\
  \textbf{Documentation \mbox{Frequency}} & 2-3 times a week (27), Weekly/biweekly (3), \mbox{Once a month (1)} \\
  \textbf{Documentation Types Created} & 
  \begin{minipage}[t]{\linewidth}\raggedright
      How-To (25), GUI Software Guides (23), Programming Tutorials (16), API Documentation (9), Product and Feature Documentation (6), Conceptual and System Design/Architecture Documentation (5), Software Deployment Guides (2), Hardware Documentation (2), 
      \mbox{Learning and Training Materials (1)}
      \vspace{2mm}
  \end{minipage} \\
  \arrayrulecolor{black}\bottomrule
  \end{tabular}
  \label{tab:participant_info}
  \vspace{-14pt}
\end{table}

We performed initial analysis on each interview transcript immediately after the interview (details in \S\ref{subsec:method_data_anlaysis}), which allowed us to iteratively develop our analysis and continue recruitment until we reached robust themes. The five review stages (RQ1) emerged within the first 15 interviews with later participants describing variations or combination of these same stages rather than fundamentally different processes. Similarly, our understanding of challenges (RQ2) within each stage stabilized when completing around 20 interviews. Later interviews provided additional examples and context for previously identified themes rather than adding significant new insights, indicating that we reached saturation on major themes~\cite{guest_how_2006, braun_saturate_2021}. We ultimately recruited 31 technical writers, henceforth referred to as $P1$-$P31$; Table~\ref{tab:participant_info} summarizes their demographic and professional information.

\subsection{Interview Protocol and Data Analysis}
\label{subsec:method_data_anlaysis}
All interviews were conducted remotely by the first author and were screen recorded, with participants encouraged to demonstrate their practices by sharing their screens when discussing specific examples. Interviews lasted between 45-70 minutes. Each participant was compensated with a gift card valued at \$45 CAD or an equivalent amount in their local currency.

Guiding questions for the interviews evolved iteratively based on emerging findings from the simultaneous data analysis, largely aligning with the interview study design outlined by \citet{charmaz_constructing_2014}. Initial interviews used broad, exploratory questions, and probed deeper depending on the participants' responses:
\begin{enumerate}[topsep=0pt]
  \item What do you consider high quality documentation? How do you achieve this? 
  \item What steps do you take to make sure that the documentation you write meets your desired quality? 
  \item Do you use any tools to improve documentation quality?
  \item What are some challenges you face to reach your desired documentation quality?
\end{enumerate}
Once we identified the general perspectives towards the documentation quality, review stages and challenges, we developed more questions as follows about specific review stages while maintaining our initial broad questions: 
\begin{enumerate}[topsep=0pt]
  \item[(2.1)] How is the documentation reviewed? Can you provide some details on the process?
  \item[(2.2)] Who are the people involved in the review?
  \item[(2.3)] Why is it necessary to perform technical review/[another review stage]? (Only if participant mentioned the stage)
\end{enumerate}
As our understanding evolved, the first author also reviewed earlier interviews to identify any themes we might have initially missed to improve validity. Once saturation on themes was reached, we began each session with the previously outlined open-ended questions, and then used these targeted questions only when participants mentioned the specific aspects, allowing us to validate our findings but keeping us open to approaches different from the findings.

All interviews were transcribed using Microsoft Teams' automated transcription; we analyzed both transcripts and screen recordings to capture the full context of participants' demonstrations. The first author conducted all 31 interviews and led the iterative analysis of the data, during which they developed a nuanced knowledge of documentation creation and reviewing practices described by our participants. Following \citet{charmaz_constructing_2014}, we began with an open coding pass to identify themes and generate preliminary codes. 

We held regular peer debriefing meetings with all authors to review codes and revise them as more interviews were conducted. Because the first author conducted all interviews and held familiarity with the data, they led each session by presenting candidate codes along with supporting examples from the transcripts. When authors interpreted a code differently, we resolved the disagreement by returning to these examples. The first author justified the code's definition against the underlying data, and we refined or relabeled the code until its grounding was clear to all authors. Rather than calculating inter-rater reliability, we used reflexivity and peer debriefing to address potential bias~\cite{mcdonald_reliability_2019}. The resulting codebook is available in the supplementary material.

\subsection{Validation} 
\begin{figure}[tbp]
  \centering
  \begin{subfigure}[t]{0.48\columnwidth}
    \centering
    \includegraphics[width=0.85\linewidth]{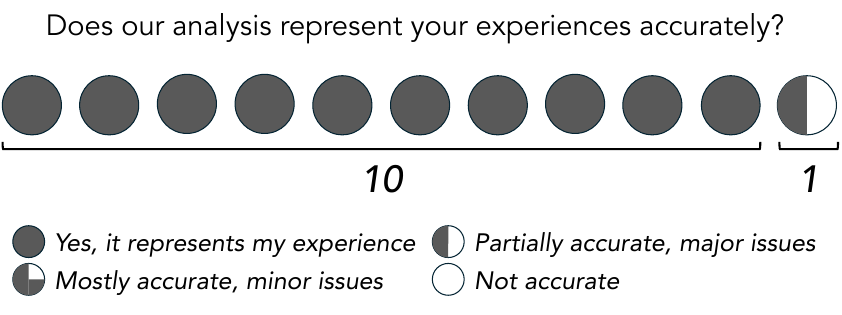}
    \caption{\hyphenpenalty=10000 Overall, do you feel our analysis accurately represents your experiences and the practices you described?}
    \label{fig:validation-a}
  \end{subfigure}
  \hfill
  \begin{subfigure}[t]{0.48\columnwidth}
    \centering
    \raisebox{-9.1pt}{\includegraphics[width=0.85\linewidth]{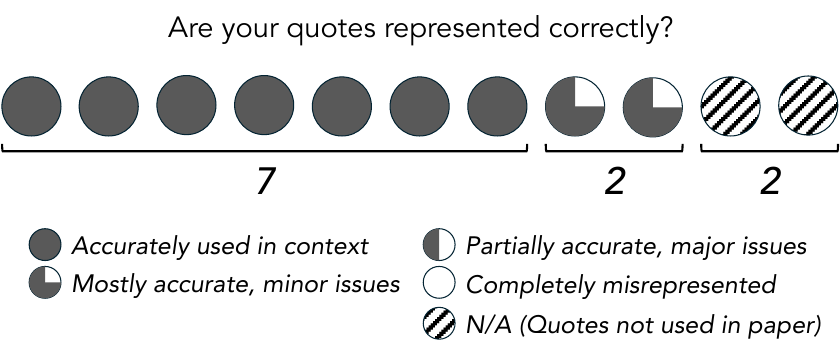}}
    \caption{If we quoted you in the paper, are your quotes represented correctly?}
    \label{fig:validation-b}
  \end{subfigure}
  \caption{Validation results from (n=11) respondents. Each circle represents one participant. Two additional participants confirmed the findings via email but are not shown.}
  \label{fig:validation}
\end{figure}

Our account of the review process was built from participants' descriptions, so we shared our findings with them and asked them to assess it and note any shortcomings. This participant validation step was informed by synthesized member checking~\cite{birt_member_2016} and validation practices in prior SE interview studies~\cite{nahar_collaboration_2022}. We shared a two-page summary of our results along with the complete draft with all 31 participants, inviting responses via questionnaire or email, in order to offer flexibility in the level of effort required. In the questionnaire, we prompted them to assess and reflect whether our analysis accurately represented their experiences, whether their quotes were used in appropriate context, and to provide detailed feedback on the review stages, design space, challenges, and coping strategies we identified. Out of the thirteen participants who responded, twelve confirmed the analysis accurately represented their experiences. \pidf{P5} indicated partial agreement (Figure~\ref{fig:validation-a}), noting that documentation quality depends primarily on the writer's independent understanding of the software, with collaborative review serving as a verification step rather than the primary mechanism. Of the nine respondents whose quotes appeared in the paper, seven confirmed their quotes were used accurately in context; two noted minor phrasing concerns (Figure~\ref{fig:validation-b}), which we clarified. Any observations made by participants during validation are integrated in the final paper and are indicated inline using a subscript next to participant ID---e.g., \pidfv{P5}. The questionnaire and the full responses are available in the supplementary material.

\subsection{Limitations} 
We describe two methodological considerations for interpreting our results. Additionally, we provide the decisions behind our research design and corresponding trade-offs~\cite{robillard_communicating_2024}, and the positionality of authors~\cite{olmosvega_practical_2023} in the Supplementary Material.

Regarding construct validity, our participants provided accounts from their own perspective. Because we study review from the writer's standpoint (\S\ref{subsec:recruitment}), all claims about reviewer behavior in this paper reflect the writer's account of that behavior. Writers may not be aware of how others in the review workflow experience the same process, including workarounds outside their direct involvement. Another analytical limitation is that our focus on identifying recurring practices may have led us to emphasize structured review stages and underrepresent informal practices that fall outside the stage boundaries we defined. We nonetheless attempted to capture these through iterative analysis and by revisiting earlier interviews as our understanding evolved.

Regarding external validity, our recruiting pool consisted of experienced participants in technical writing roles and active in professional communities like Write The Docs and LinkedIn. This means we primarily capture perspectives of experienced technical writers from companies that invest in dedicated technical writing roles and have more systematic review processes. While this provides a detailed understanding of review processes, it may miss perspectives from novice technical writers, other roles like developers or program managers, and informal documentation review contexts in organizations with less systematic practices.
 
\section{RQ1: How to Achieve Documentation Quality}
\label{sec:doc_quality}
Since practitioners cannot directly measure whether users achieve their goals through documentation, they instead evaluate documentation based on quality attributes they believe contribute to user success. Therefore, understanding their perception of ``quality'' is necessary to contextualize our findings. Our participants repeatedly referenced seven quality concerns during their discussions of review, which we triangulate into quality attributes discussed in prior software engineering research. We organize these attributes into three categories: \emph{Content Quality} \content, which encompasses the factual correctness and comprehensiveness of information; \emph{Presentation Quality} \presentation, which relates to how information is structured and communicated; and \emph{User Experience Quality} \userexp, which concerns how users discover and successfully interact with documentation. These categories, along with representative quotes, are shown in Table~\ref{tab:quality_aspects}. We consider these categories to structure our findings because reviewers assess related attributes simultaneously rather than one at a time. For example, developers evaluate accuracy and completeness together during technical review, rather than addressing them separately. Assessing these attributes reliably, however, presents challenges. 

In \S\ref{sec:doc_quality} and \S\ref{sec:challenges_review}, we indicate the number of participants who reported each practice or challenge in parentheses. These counts serve only to report our study rather than signal the relevance of a stage, since we did not ask earlier participants about stages we identified later in our interviews.

\subsection{What compels documentation review?}
From participants' input, we identify two main reasons why a systematic review process is considered essential for achieving high-quality documentation.

\begin{table}[t]
\small
\caption{Documentation quality aspects organized into Content \content, Presentation \presentation, and User Experience \userexp. We note the number of participants who mentioned each aspect in parentheses.}
\rowcolors{2}{white}{gray!15}
\begin{tabular}{L{0.2\columnwidth}L{0.19\columnwidth}L{0.57\columnwidth}}
\toprule
\textbf{Quality \mbox{Categories}} & \textbf{Quality \mbox{Aspects}}\dag & \textbf{Representative Quote} \\
\midrule
Content \content & Accuracy (16) & \qpt{...quality to me always starts with accuracy... because if it's not accurate, it's not useful. Period.}{P5} \\
 & Completeness (16) & \qpt{...as complete as possible, having every corner of the product covered so that people can go to the doc site and find any information.}{P13} \\
\arrayrulecolor{black!50}\midrule
Presentation \presentation & Language (20) & \qpt{...quality means well written, one voice, with a style guide, concise.}{P8} \\
 & Structure (17) & \qpt{...if the structure is good, the content writing is the easy part...}{P20} \\
 & Consistency (14) & \qpt{...glossary at the end of every single document so that when somebody encounters that term, they can go to the glossary and also to standardize the term across the company.}{P25} \\
\midrule
User \mbox{Experience \userexp} & Findability \textsuperscript{\ddag} (8) & \qpt{...qualities like the docs are findable, searchable, richly linked.}{P5} \\
 & Usability (13) & \qpt{[we don't want docs to be] ugly to look at, or slow, or not support dark mode... there are lots of quality of life things.}{P14} \\
\arrayrulecolor{black}\bottomrule
\rowcolor{white} \multicolumn{3}{p{0.98\columnwidth}}{\dag We use the definitions for quality aspects from \citet{zhi_cost_2015, treude_beyond_2020,aghajani_software_2019, aghajani_software_2020, tang_evaluating_2023}.} \\
\rowcolor{white} \multicolumn{3}{p{0.98\columnwidth}}{\ddag\  Instead of using \textit{Accessibility}, used in both \citet{zhi_cost_2015} and \citet{aghajani_software_2020}, we consider \textit{Findability} better describes this quality aspect; and it avoids confusion with a broader research topic of Accessibility.}
\end{tabular}
\label{tab:quality_aspects}
\vspace{-12pt}
\end{table}

\subsubsection{Users Do Not Provide Clear Indicators of Documentation Quality \pc{17}}
\label{subsubsec:unreliable_user_feedback}
Participants shared the struggle to obtain feedback from users despite implementing several feedback collection channels, including implicit mechanisms like web traffic analytics and explicit mechanisms like voting and comment sections on the documentation site. However, the data collected through these channels always needs further interpretation, as illustrated by \pidf{P3}, \qpwi{You can put some analytics and see what people are looking at. However, that doesn't necessarily tell you whether it's successful or not... If somebody's looking at a section frequently, it might mean we have a problem in the product itself.} Participants reported that even when users provide feedback, it is not on quality attributes, possibly since users lack the specialized knowledge to evaluate technical accuracy, or care about the stylistic requirements, as \pidf{P6} reasoned, \qpwi{I don't think the regular end user really cares if you use consistent formatting or consistent wording.} 

\subsubsection{Individual Practitioner Perspectives are Limited \pc{21}}
\label{subsubsec:individual_practitioner_perspective}
Participants reported that different reviewers would judge the same document differently as their criteria for evaluation is influenced by their professional background, preferences, and individual standards. Moreover, a single reviewer may not fully understand or anticipate all potential audience needs and knowledge gaps, leading to documentation that serves some users but fails others. To counter this, collaborating with experts can reveal incorrect assumptions, identify blind spots, and include different perspectives. \pidf{P4} illustrated this point, \qpwi{[What I need is] \textbf{a second set of eyes.} Someone who has a different perspective than mine, and ideally, this person also understands the audience that the content is aimed at. They would be able to tell me if the content would meet the audience's expectations or confirm that my assumptions about the audience's expectations make sense.} Beyond catching any individual oversights, collaboration also helps to resolve conflicting viewpoints about documentation scope and content. These conflicts arise because reviewers bring different assumptions about what the documentation is for. \citet{liu_articulation_2023} term such gaps as \emph{intersubjectivity disjunctures}, which are the points where collaborators' differing assumptions must be worked out for work to proceed. \pidf{P12} described how discussions during review help achieve this, \qpwi{An engineer's perspective of the feature is going to be somewhat different than a PM's perspective of a feature... Getting at least PM and engineering to both agree also helps with questions down the line of `Why didn't you include this?' or `That feature's not actually generally available, it's still beta.' Getting that consensus on features and what should go to our customers is very helpful.}

\subsection{Documentation Review Process}
\label{subsec:doc_review_process}
\subsubsection{Overview}
\label{subsubsec:doc_review_overview}
Documentation quality rests primarily on the technical writer's expertise and their independent exploration of the software. The review process addresses blind spots inherent to any individual perspective. In addition, because the information a document needs to contain is spread across several practitioners, technical writers must recruit the right reviewers, direct relevant parts of the document to them, and reconcile their feedback into a single artifact. This effort to align contributions from different people so that a collective task proceeds coherently is called \emph{articulation work} in the CSCW literature~\cite{strauss_articulation_1988, corbin_articulation_1993}. We show how documentation review operates as articulation work through by identifying five stages of reviews that documentation can undergo (depicted in Figure \ref{fig:process_overview}), with each review stage performed by different practitioners and fulfilling distinct quality categories. We observed three main review stages that take place after drafting the documentation: \emph{self review}, \emph{technical review}, and \emph{editorial review}. These stages occur most frequently and are conducted internally to the team or organization. How these stages occur varies; any stage can happen first and sometimes multiple times in an iterative manner. \pidfv{P3} confirmed that stages also overlap, for example when technical and editorial feedback occur within the same pull request conversation. In addition, we observed two stages, \emph{play testing} and \emph{post publication feedback}, in which documentation receives further feedback, primarily geared towards ensuring completeness and verifying that the content is useful to the audience. We describe the \textbf{practitioners involved} and the \textbf{quality categories} they address in each stage; together these show the \emph{arrangements}~\cite{corbin_articulation_1993} that structure each stage, meaning who contributes and against which standards their work is judged.

We observed variations in how different organizations adopt these review stages. Participants from organizations that prioritized faster software releases described shorter review processes, while participants from regulated industries like aviation and military that require strict compliance described strict review processes, formal change requests and approval workflows. Participants who were the only writers in the team reported publishing documentation without editorial oversight, while larger teams consult specialized experts like lawyers or marketing staff, sometimes outside standard processes---e.g., \qp{Depending on the purpose of the document, it needs to have a little bit of marketing lingo in it, and that's when we bring in the marketing team [to provide feedback].}{P17} This variation suggests that organizations adopt the required review stages based on their contextual needs rather than as a prescribed mechanism.

\subsubsection{Self Review \pc{27}.}
\label{subsubsec:self_review}
\subparagraph{\textbf{Practitioners Involved.}} We define self review as technical writers~(\utw) evaluating their own work. Most of our participants were technical writers who regularly performed self reviews, while three participants who reported their roles as manager and developer described supporting this process. Unlike the stages that follow, self review is conducted largely alone and therefore does not involve significant coordination effort. It is instead the \emph{primary work}~\cite{schmidt_taking_1992} of producing the document, on which the collaborative stages and their articulation work are built.

\subparagraph{\textbf{Content Quality \content.}}
Participants reported evaluating content quality by testing documentation against the software, cross-referencing with reliable information sources, and seeking expert validation. For testing against the product, participants execute documented steps within the software. This hands-on approach varies by documentation type; GUI documentation requires direct verification within the software interface, API documentation involves testing calls and verifying responses. \pidf{P4} explained how they test code examples within the documentation, \qpwi{In most cases I can just run the code directly on the system to make sure it's working. If it gives the right output and no errors, then it's obviously the correct code.} Writers with extensive product knowledge can conduct this testing independently, while those with less expertise supplement their testing with input from Subject Matter Experts (SMEs). Participants also cross-reference their work against reliable sources like JIRA tickets and recorded meetings, and validate their understanding with SMEs and developers for complex topics as illustrated by \pidf{P1}, \qpwi{If it's a complicated topic, I put in a few ideas of what I think it is and then send it to either the product manager or the dev to say, `Hey, am I on the right track?' If they confirm yes, then I go deeper.}

\subparagraph{\textbf{Presentation Quality \presentation.}}
Participants reported that they review presentation quality by using their internalized knowledge and by checking organizational guidelines when needed. Experienced writers rely on knowledge developed over years of practice to identify presentation issues, \qp{I used to be a copy editor as well, and I am pretty experienced and good at finding grammatical errors and spelling errors... They just kind of stand out to me, I think. I can spot a comma splice a mile away.}{P27}  Participants consult external style guides when needed to develop their knowledge about and stay consistent with expected presentation quality. When unsure about specific elements, they consult industry standard style guides like the IBM guide or company specific standards, and reference centralized terminology resources and templates to maintain consistency across documentation. \pidf{P7} illustrated this, \qpwi{We follow the IBM style guide, and on top of that, we have a supplementary style guide specific to Red Hat. This lists how we handle Red Hat product names and other things.}

\subsubsection{Technical Review \pc{28}.}
\label{subsubsec:tech_review}
\subparagraph{\textbf{Practitioners Involved.}} Participants reported that technical review focuses on evaluating documentation for technical accuracy and completeness. They identified the developers~(\udev) of the software features and SMEs as the primary reviewers in this stage due to their deep technical understanding, as described by \pidf{P5}, \qpwi{The technical review is always going to be from the engineer that wrote the feature. If I'm lucky, the product manager will take a look, but usually they just say, `Yeah, sure, fine, looks good.'} 

\subparagraph{\textbf{Content Quality \content.}}
Participants reported that during technical review, developers test documented procedures by running API endpoints, building documentation environments, and creating tests to verify documented functionality. They described developers as identifying missing prerequisites and incomplete instructions, and as validating visual elements like screenshots and code samples for technical correctness. Technical review also serves as a safeguard against any editorial improvements accidentally removing important technical details. \pidf{P11} explained, \qpwi{We try to understand if we have not lost the technical accuracy of the feature... there might be things like known issues with the feature that they have written in that was between the lines that we could have edited out because we thought it was language and not an actual feature.} Additionally, review sessions help participants identify implicit knowledge that never appeared in development tickets or feature specifications. Participants described that SMEs share contextual information during review conversations when they notice gaps between the documentation and how the system actually works. Writers then probe these insights by asking questions and following up on casual comments since experts often possess critical information they consider obvious but users would find essential. In some instances, developers and SMEs insist on documenting every feature to showcase their work, or conversely, dismiss information they consider obvious but users find essential, an intersubjectivity disjuncture~\cite{liu_articulation_2023} between documenting the system and documenting the user's path through it. In these instances, participants described pushing back against developers, as \pidf{P21} noted, \qpwi{The engineers care very deeply about how something works under the hood. They want to explain every single button they made and every single way that you can possibly configure something, whereas I care very deeply about the workflow for the user. There's always this negotiation of like, do we really need to document this?} These negotiations, through which collaborators persuade one another and settle on what the documentation should be, are an instance of articulation \emph{working out} or reconciling the competing positions~\cite{corbin_articulation_1993}.

\subsubsection{Editorial Review \pc{28}.}
\label{subsubsec:edit_review}
\subparagraph{\textbf{Practitioners Involved.}} Editorial review focuses on evaluating the language, structure and user experience of the documentation. Most participants described relying on fellow technical writers~(\utw) who understand organizational style guidelines and can evaluate documentation from a user perspective. In organizations without dedicated editors, peer technical writers fulfill this editorial role. Participants also described involving product owners and managers~(\upm) to bring in brand and legal considerations, \qp{Product owners have a set brand that they're looking for. They have words they want to avoid, like [this word], which insinuates that we have this capability... Readability is also tied into liability, compliance, legal etc, and we need to make sure we're not either overpromising or under delivering.}{P30}

\subparagraph{\textbf{Presentation Quality \presentation}.}
Participants described that documentation needs to maintain a `singular organizational voice' despite having multiple contributors, which is achieved by maintaining a style guide specific to the organization. This style guide acts as a \emph{common information space}~\cite{schmidt_taking_1992}. However, such shared resources never carry a single fixed meaning; in practice, writers reading the same guide interpret and apply it differently. During editorial review, these differing interpretations are reconciled as reviewers check the documentation's adherence to organizational style guides and provide feedback on structural elements like bullet points, headings, and language quality. This includes ensuring content matches the appropriate tone for different document types, for example, straightforward and clear for technical documentation and more persuasive language for marketing materials. This consistency becomes especially important when technical writers have different language backgrounds. \pidf{P30} described that editorial review provides language support that allows writers to focus on technical accuracy while reviewers handle style consistency, \qpwi{A lot of my team is ESL (English as Second Language). They're kind of reluctant to write because I think they're worried that I'm going to be prescriptive, but I say I don't really care as long as the information is correct, I will make it look good. I also encourage them to use any writing assistants, if they have access.} \pidfv{P5} distinguished between style and design aspects of presentation quality. Style aspects like tone, terminology, and grammar can be partially automated through linters (\S\ref{subsubsec:proactive_timing}), while reviewers assess visual design by previewing rendered documentation pages for ability to skim and adherence to web writing conventions.

\subparagraph{\textbf{User Experience Quality \userexp.}}
Participants reported that as they focus on creating documentation, they often overlook whether users can actually discover and apply the information successfully. We identified three common user experience quality concerns addressed during editorial review. First, basic issues like broken links can prevent users from reaching content regardless of its quality, so reviewers check all links within submitted documentation pages by accessing them. Second, gaps like jumping between ideas without clear connections can confuse users trying to follow documentation, which happens because technical writers become too close to their material to recognize when they have created these narrative breaks. \pidf{P22} described how reviewers help fix such narrative errors during editorial review, \qpwi{I'm looking for any plot holes, cause it's essentially a story right? If we're jumping from chapter 2 to chapter 8, with nothing in between, then I'm gonna lose the thread of what I'm supposed to be doing.} Finally, writers often structure content around their own technical understanding rather than considering users with different expertise levels. Reviewers from different domains help identify these assumptions and ensure that the content works for users across various technical backgrounds.

\subsubsection{Play Testing \pc{13}.}
\label{subsubsec:play_testing}
\subparagraph{\textbf{Practitioners Involved.}} 
Play testing requires reviewers who can approach the documentation with fresh eyes to conduct end to end verification. In most cases, participants reported involving quality assurance teams~(\uqa) who possess product knowledge but were not involved in contributing to the documentation. \pidf{P18} explained that the key qualification for play testing is this outsider perspective, \qpwi{I had people from my team pretend that they were a user with no knowledge of anything and go through a document and walk me through it live, through their thought process and where things didn't make sense.}

\subparagraph{\textbf{Content Quality \content.}}
Participants reported that reviewers verify content quality during play testing by evaluating documentation from a user perspective, in contrast to technical review which relies on expert knowledge. They conduct hands-on verification by following documented steps systematically on the production software to confirm that instructions work as described and documentation accurately reflects the software functionality. In addition, the focus is on identifying gaps that writers might miss due to their familiarity with the system, what participants termed as the `curse of knowledge,' where experts incorrectly assume readers possess the required background. \pidf{P5} explained, \qpwi{I might be looking for things like did I skip a prerequisite? Did I jump in complexity too quickly?... Play testing can catch if you got sloppy or skipped some stuff. It checks for that curse of knowledge in a tutorial, did I forget to explain something because I understand it, but it's not known to someone who's new?} Fewer participants reported play testing as part of their standard review workflow. \pidfv{P3} noted that it tends to be informal and opportunistic, dependent on reviewer bandwidth from teams like QA or customer success. \pidfv{P5} added that it occurs primarily for tutorials and is rare for other documentation types.

\subsubsection{Post Publication Feedback \pc{18}.}
\label{subsubsec:post_pub}
\subparagraph{\textbf{Practitioners Involved.}} Post publication feedback involves actual product users~(\uext) or roles like solution architects who use the documentation to deploy in production environments. Writers rarely reach the actual users of the product directly and instead depend on proxy measures like support tickets or site analytics. These measures were not built to capture documentation quality, so quality issues surface in them only indirectly~\cite{star_layers_1999}. Reading these signals therefore requires interpretation, and writers rely on customer support staff, who see the user's actual struggle, to sort which reported problems are documentation gaps.

\subparagraph{\textbf{Content Quality \content.}}
This feedback primarily reveals two types of content gaps: missing implementation steps when customer environments differ from development settings, and undocumented use cases when users use the software in unexpected ways that developers did not anticipate. As \pidf{P23} noted, \qpwi{SMEs might not catch certain scenarios that happen in customer environments. Their testing in development environments or unit tests might not cover all customer scenarios. When issues are reported, it might lead to adding notes in the documentation or revising content.} This feedback helps the technical writers prioritize documentation updates based on what information delivers practical value in real world contexts. They might make minor content adjustments for small fixes, or conduct major revisions leading to new documentation versions.

\subparagraph{\textbf{User Experience Quality \userexp.}}
Participants mentioned that users face challenges with finding the information due to unavailability in search because of poor organization of content or insufficient metadata within documentation. Participants reported that findability issues are visible in auto-generated technical documentation like API references, which are not systematically reviewed since they are directly generated from code, and therefore only way to discover issues is through complaints from users. Furthermore, different user groups require distinct presentation approaches even when documenting the same functionality which can be restructured based on feedback from the audience, as described by \pidf{P20}, \qpwi{A document that would be perfect size for a software engineer, [but] too long for a firefighter. I had to rewrite the whole thing, and to me it looked like `Oh, this is for little kids' because it was only three lines, table, another three lines, table... but that's what the firefighter needed.} \pidfv{P5} pointed out that the docstrings accompanying auto-generated API specs are often reviewed by writers, so findability challenges in these cases reflect tool limitations in how the generated documentation is presented to users rather than a lack of review. As \pidf{P23} noted, \qpwi{We've received feedback from customers saying it's slow and difficult to search API documentation. We use OpenAPI to generate API documentation from the source code but if the developer doesn't add a clear description, the documentation doesn't convey the right meaning.}
\subsection{Tools Used in Review Process}
\label{subsec:tools}
\begin{figure}
    \centering
    \includegraphics[width=\linewidth]{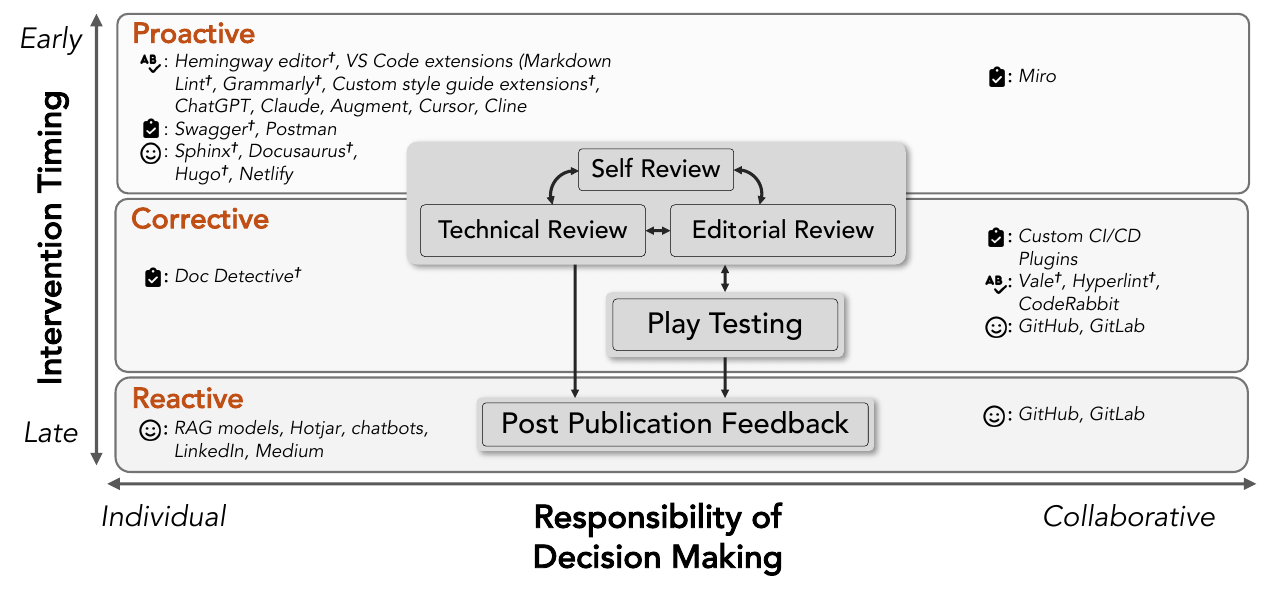}
    \caption{We organize the tools mentioned by participants in our interviews along two dimensions of Intervention Timing (\emph{when} tools are used in the review process) and Responsibility of Decision Making (\emph{who} makes documentation-related decisions with them). We indicate the quality category a tool supports: Content \content, Presentation \presentation, and User Experience \userexp. Tools created specifically for documentation are marked with a \dag\, sign; general software engineering tools adapted for documentation work are unmarked. This mapping reveals how documentation work relies on many tools \emph{not} designed for documentation, indicating opportunities for bespoke tools.} 
    \label{fig:tools_overview}
    \Description{Tools identified by the participants plotted on a grid. The vertical axis is Intervention Timing, running from Proactive at the top through Corrective to Reactive at the bottom. The horizontal axis is Responsibility of Decision Making, running from Individual on the left to Collaborative on the right. Each tool is tagged with the quality category it supports.}
    \vspace{-12pt}
\end{figure}

The review process is instrumented with various tools. We characterize the tool landscape as a design space along two dimensions, \textit{Intervention Timing} and \textit{Responsibility of Decision Making}, shown in Fig.~\ref{fig:tools_overview}.

\textbf{Intervention Timing (T)} describes when quality interventions occur during the documentation lifecycle: \textit{Proactive} involves quality mechanisms that operate during drafting and self review to prevent issues from occurring, such as providing clear expectations regarding style and language standards; \textit{Corrective} refers to mechanisms to identify and correct issues before publication, as well as the iterative self review that occurs as writers incorporate feedback from those stages; finally, \textit{Reactive} involves mechanisms that improve quality after publication based on real-world usage feedback, such as ways to investigate documentation usage and prioritize updates. Self review spans both proactive and corrective timing, since writers review their own drafts as they write and again as they incorporate reviewer feedback. This shows that writers move between the primary work of writing and the articulation work of review throughout the process. \textbf{Responsibility of Decision Making (DM)} characterizes whether quality-related decisions rest with a single \textit{Individual} or are resolved through a \textit{Collaborative} process. Individual decision making involves a single person evaluating quality and deciding, even where they consult others. Collaborative decision making involves multiple people evaluating documentation together through discussion and negotiation, which brings in broader expertise but adds the coordination overhead of articulation work.

Our dimensions serve as one analytical lens for researchers and tool designers to identify gaps and opportunities across the review process. Practitioners may conceptualize their tooling differently, depending on their practical needs; \pidfv{P5} described thinking about tools in terms of what objective they serve and where they integrate into the workflow (e.g., editor extensions, CI/CD actions, scripts that run periodically). 

\subsubsection{Proactive \pc{25}.}
\label{subsubsec:proactive_timing}
Participants mentioned that they tried to prevent certain issues from occurring, and discussed the tools while creating documentation and performing early reviews.

\noindent \textbf{Content Quality \content.} Several participants reported that they extensively test the documentation during and after it is completed using APIs, test databases, and staging environments. They validate code snippets and workflows using tools like Postman with test databases to verify API documentation accuracy. Some participants reported using collaborative tools like Miro for real-time collaborative feedback, especially on complex diagrams during creation. Participants were starting to increasingly rely on LLMs for content validation, such as replacing traditional tools like JSONLint for checking JSON, or prompting LLM to validate the code snippets and documentation. \pidf{P6} described a personal experiment of using LLMs to review content by \qp{[asking] the LLM to read documentation and review it as an LLM, asking if there's anything in the documentation that an LLM would find confusing... Just [prompting], `You are an LLM reading this text and answering as an LLM. What do you find confusing?'}{P6}

\noindent \textbf{Presentation Quality \presentation.} Several participants relied heavily on editor based tools for receiving immediate feedback. VS Code is considered a standard tool in the technical writing field, and several participants reported extensions like Markdown Lint to get immediate feedback as they draft the documentation. The main benefit of using these extensions is that they provide visual feedback in real-time by underlining errors. To improve language, participants reported using Grammarly~\cite{grammarly_grammarly_2025} and Hemingway editor to identify critical grammar issues like passive voice and improve clarity for non-native English speakers. Participants reported enforcing organization-specific style guides and terminology requirements by adopting custom developed extensions. \pidf{P23} explains the impact of these tools, \qpwi{Before implementing linting and the style guide, it was difficult because we all had different writing styles. Now it's easier and faster to review because we focus on whether the concept is clear and understandable.} 

Participants reported using LLMs like ChatGPT for basic corrections (spelling, grammar, and structure), style improvements (sentence clarity, rephrasing, conciseness, and voice conversion), and specialized tasks (readability adjustment, word selection, and custom style enforcement). Participants reported uploading the style guides directly into ChatGPT to receive automated feedback or create custom style settings in LLMs like Claude where this feature is available to maintain uniform writing styles. \pidfv{P5} highlighted an emerging shift toward coding agents like Cursor, Cline, and Augment that integrate LLM capabilities directly into the editor environment, providing documentation assistance within the same workflow. They described using these agents for formulaic tasks such as marking features as deprecated or in beta across multiple pages, where the agent propagates the same pattern across the documentation rather than the writer doing it manually.

\noindent \textbf{User Experience Quality \userexp.} 
To address the primary user experience issues with the documentation sites, several participants reported using documentation site frameworks like Sphinx or Docusaurus, which support local builds, or staging on web hosting sites like Netlify, to visually inspect these aspects before publication. Several plugins are integrated into the CI/CD pipeline used to build the website, which checks issues like broken links, missing image paths, and faulty URLs before documentation is pushed to the repositories. \pidf{P8} mentioned the need to automate checking the validity of the links \qpwi{to make sure we know about broken links... Because we use markdown files, whenever we remove a file, we have to put a redirect. We have to do that because otherwise you get a 404 page, which is bad for Search Engine Optimization (SEO) [which] is big consideration as well.}

\subsubsection{Corrective \pc{14}.}
During the technical and editorial review stages, participants reported using tools to assist in evaluating both content and presentation. 

\noindent \textbf{Content Quality \content.} Participants reported using automated documentation testing tools like Doc Detective~\cite{manny_docdetective_2025} that create parallel tests to validate user tasks and commands. Such tools, during pull request review time, execute documented procedures to verify that instructions actually work as described, and catch functional errors that manual review might miss. Most participants mentioned integrating custom tooling and plugins in CI/CD pipelines to test the functionality of documented screenshots and API endpoints during the review process, ensuring that instructions remain accurate before publication. Participants also envisioned how LLMs can be used on the repository for more detailed feedback, for example, \qp{One of the biggest problems we face in terms of maintaining accuracy is we identify docs A, B and C need to be changed, but there's one sentence buried halfway down on Doc D that is no longer accurate... [the tool must] pull the five most semantically similar guides from the rest of the doc set and compare to make sure it is accurate.}{P14} \pidfv{P5} added that this vision is already materializing through LLM-based review agents like CodeRabbit that analyze pull requests automatically, complementing the rule-based CI/CD checks described above with feedback that accounts for documentation semantics.

\noindent \textbf{Presentation Quality \presentation.} Participants reported using CI/CD plugins integrated with their version control systems to provide an avenue for discussion and enforce presentation quality collaboratively. Participants extensively licensed software like Vale~\cite{errataai_vale_2025} and Hyperlint~\cite{hyperlint_hyperlint_2025} for programmatic analysis of pull requests to detect style violations and ensure quality standards, with the build pipeline blocking merges until issues are resolved. Organizations create custom Vale rules based on specific style guides like IBM standards, making writers responsible for compliance on every pull request. \pidf{P23} describes how this mechanism prevents problematic content from being published, \qpwi{If I overlook an error and try to publish it in Cloudflare, the process catches it and doesn't let me publish the doc. It says `build failed because these validations failed.'} 

\subsubsection{Reactive \pc{18}.}
\label{subsubsec:reactive_timing}
Participants described tools used for collecting feedback and the changing consumption of documentation. 

\noindent \textbf{User Experience Quality \userexp.} Participants reported employing multiple channels like votes and comments for receiving user feedback and supplementing this information with support ticket counts for contextual information regarding issues. \pidfv{P3} noted that teams also rely on behavioral analytics tools such as heatmaps and search query analysis to identify content gaps and understand how users navigate documentation sites, supplementing explicit feedback channels with implicit usage data. Most participants use version control platforms like GitHub or GitLab to host their documentation, which also facilitates users to report issues and add comments directly within documentation repositories. \pidfv{P5} clarified that few organizations open-source their documentation, so this channel primarily captures internal user feedback; external feedback more commonly arrives through on-page comments, community forums, or platforms like Reddit. Finally, some participants also discussed how users are increasingly using LLMs to consume the documentation, and how their goal is to create documentation that can be used to train the LLMs better. Several participants had already started using advanced search functionality powered by custom retrieval-augmented generation models to generate summaries based on user queries, or chatbots to allow users to receive tailored answers on their documentation websites. 
\section{RQ2: Challenges During Review}
\label{sec:challenges_review}
In the previous section, we described the articulation work writers perform, such as recruiting reviewers, directing parts of the document to them, and reconciling the received feedback, all of which is an overhead that distributed work like review imposes~\cite{schmidt_taking_1992}. This overhead is not usually recognized as separate from the work of producing documentation, and surfaces only in the challenges we report in this section, where it broke down enough for participants to name it~\cite{strauss_articulation_1988}. When reporting each challenge we indicate how it corresponds to the design space introduced in \S\ref{subsec:tools} as in \textbf{\hlc{(T, DM)}} (\textbf{\hlc{*}} indicates overarching concerns along the entire dimension).

\subsection{Process Challenges}
\label{subsec:process_challenges}
\subsubsection{Poor Planning of Documentation Cycles \pc{21}} \textbf{\hlc{(Proactive\slash Corrective, *)}}
\label{subsubsec:process_challenge_planning}
Participants reported that documentation typically lags behind engineering sprints, forcing writers to perpetually chase development within an \emph{arrangement}~\cite{corbin_articulation_1993} whose terms are set by the development teams. This results in short documentation windows; writers have little time to document features after they reach staging before being pushed to production. Consequently, participants find themselves documenting incomplete and untested features, and since they cannot verify the accuracy, the content quality is reduced, as illustrated by \pidf{P30}, \qpwi{You will get a request to document the deliverable [after] around 90\% of development [is completed], and then you have one week to write 86 pages on a integration guide. You're like, that's not gonna happen, and they're like, well, it needs to happen. Then you say okay, but that usually means there are knowledge gaps.} The deadlines remain inflexible due to external constraints like product launches, marketing campaigns, and regulatory requirements, yet additional documentation requests continue without adjusting the timeline. For example, \qp{Some SME will just pop in when they have time and say, `Oh, we need to add an entire section on [XYZ],' which adds maybe 30 pages, and the deadline doesn't change because we still have to roll it out. So that's a big challenge, people not understanding that there's time needed if you say add a section... It isn't something that's instantaneous.}{P21}

\subsubsection{Competing Priorities \pc{17}} \textbf{\hlc{(Proactive\slash Corrective, Collaborative)}}
\label{subsubsec:process_challenge_prioritizing}
Participants mentioned that reviewers prioritize their core responsibilities over documentation review, and writers hold no authority to compel them to choose otherwise. The writer's \emph{stance}~\cite{corbin_articulation_1993} in the review process is weak, since they depend on people whose working conditions they cannot influence, and persuading reviewers is the only option available to them. For example, \pidf{P5} described the challenge of persuading busy developers to review documentation, \qpwi{Reviews are not easy to get because people don't want to do stuff that is not their primary goal. You're just like, `Hey, I know you're busy writing code or something, but in order to publish your code to other people, you have to look at this document, please.'} Participants also reported that even when reviews do happen, they are done based on the convenience of the reviewer, which can result in last minute reviews. In some cases, participants mentioned that they also face resistance from colleagues who view them as creating additional burden rather than adding value. \pidf{P22} describes the interpersonal friction, \qpwi{I'm seen as the bad guy for saying, `Hey, this needs more attention' because I'm the bearer of bad news. It's challenging to not only get somebody to appreciate what I do, but also be excited to collaborate, cause a lot of the times it's a negative perception---`This person's creating more work for me'---but no, I just want to get it right.}

\subsubsection{Expertise and Communication Gaps \pc{19}} \textbf{\hlc{(Proactive\slash Corrective, Collaborative)}}
\label{subsubsec:process_challenge_expertise}
Participants reported that senior technical writers can sometimes possess deeper product knowledge than their reviewers, especially for those who have, over the years, gathered information from multiple experts across the board, like PMs, developers, and user support teams, and spent considerable time synthesizing these insights. Such an expertise mismatch between the technical writer and reviewer can result in the review process acting as a formality, as illustrated by \pidf{P6}, \qpwi{For the senior writer---she's been here 8 years---she works on some areas of the product that I don't really know myself, so for stuff she works on, I review it just for typos... For senior writers, my review is more like a rubber stamp. For junior writers, I read everything.} This sentiment was further echoed by \pidfv{P3} during validation. In these cases, the review functions mainly to satisfy the arrangement~\cite{corbin_articulation_1993} that a document must be reviewed before it ships. \pidfv{P5} went further, arguing that documentation quality rests primarily on the writer's own understanding of the software, and that collaborative review verifies that understanding rather than producing quality itself. Our findings support this for senior writers, whose work is approved without substantive engagement, while junior writers receive the scrutiny that makes review a quality mechanism.

Knowledge and communication gaps also lead reviewers to focus on minor details instead of critical issues, or to approve documentation without reading it. Furthermore, participants described that reviewers could lack knowledge about the research the technical writers have conducted, the documentation's purpose, and the information needs of the audience. In addition, language and cultural differences among team members can lead to debates over the best way to communicate the information. \pidf{P14} described this superficial engagement, \qpwi{[Reviewers] will either say nothing, they'll just ignore you, they'll say `LGTM' [Looks Good To Me] even though they didn't read it. I've seen them open a doc, they read the title, they read the first line, and then they read heading and subheadings, and by the time they [reach] the bottom, they're not reading anything at all.} \pidfv{P3} also reflected that reviewers disengage from long documents or frequent review requests, reading carefully at the start but skimming or stopping before reaching later sections.

\subsection{Tooling Challenges}
\label{subsec:tooling_challenges}
\subsubsection{Lack of Dedicated Tooling for Documentation \pc{21}} \textbf{\hlc{(*, *)}}
\label{subsubsec:tool_challenge_dedicated}
Participants reported that they have to adopt tools designed for developers, despite documentation workflows involving technical writers, product managers, and other reviewers with varied technical expertise. The steep learning curves of these developer-focused tools exclude non-technical roles from the review process, so writers work around them~\cite{gasser_integration_1986} by sharing documentation through Google Docs or exporting to PDF for feedback. These workarounds keep the review moving without repairing the tools, which stay designed for code. \pidfv{P5} noted that Google Docs is particularly common because its collaborative commenting resembles GitHub reviews; however, any changes must then be manually ported back to source files, adding friction to the workflow. \qp{When I'm reviewing with less technical stakeholders like PMs, they don't like the PR process. Giving them a code based PR to review can be difficult, and they don't always understand exactly how what they're looking at ends up being represented in the docs.}{P2} 

These tools also prevent reviewers from effectively evaluating the user experience because they present textual diffs and flattened files instead of showing the visual layout of the final published website. \pidf{P21} described this need, \qpwi{[Reviewers] are still just looking at the back end version of it. Ideally you'd be able to go onto [production] website and highlight something and comment on it, or there would be another `review' state [of] the website... It would still be the exact same website, but people with access through logins could use suggestion and comment mode.} Additionally, participants reported that version control systems cannot distinguish between meaningful content changes and automatic formatting changes. \pidf{P22} described tracking changes manually, \qpwi{When you have a lot of people tracking those changes, after a while it gets very hard to tell what the changes were. So when somebody sends something to me with their proposed edits, I save a copy of that and then change the file name to something like `Tech Edit Version' so that there's a clear distinction between what I received and what I produce.}

\subsubsection{Tool Maintenance Overhead \pc{16}} \textbf{\hlc{(*, *)}}
\label{subsubsec:tool_challenge_maintenance}
Despite seeking and championing better documentation and review solutions, participants find that the responsibility of maintaining them eventually falls on individual members. This maintenance work includes ongoing updates, configuration, and troubleshooting, which competes with their documentation responsibilities like writing and reviewing. Moreover, this sort of ownership creates a single point of failure where the tool's effectiveness depends entirely on one person's continued involvement and availability. To avoid this dependency, participants choose commercial solutions with dedicated support teams over custom implementations. \pidf{P6} described this tradeoff while choosing a commercial software solution for reviewing, \qpwi{Long term, if we implemented something like docs as code, I feel like it would depend too much on me, so it would not be maintainable... With [commercial solutions] you have a support team. If I quit my job and they don't know how to do it, they can call the support line and they will have explanations.} LLMs present similar maintenance challenges, as they require participants to develop and refine prompts to match organization specific style guides and review criteria, as both LLM capabilities and organizational needs evolve. Furthermore, effective use of LLMs requires participants to take responsibility for staying current with best practices and troubleshooting when LLM outputs fall short of review standards.

\subsubsection{Unreliability of LLM Generated Feedback \pc{25}} \textbf{\hlc{(Proactive, Individual)}}
\label{subsubsec:tool_challenge_reliability}
Participants reported that the LLMs' technical limitations and their organization's security policies prevented them from providing the full documentation and its supporting information needed for a review. To fit the context window, they broke longer documentation into smaller sections, and converted content to plain text because LLMs struggled to interpret tables, XML markup, and hyperlinks. \pidf{P1} described this, \qpwi{When I put in a copy of this content, [LLM] doesn't know where in the documentation [the content] is. I might have explained something further up in an early introduction, and [LLM] doesn't know what I've hyperlinked. So it brings all that out [as feedback] and says that a person might not know what the source is, but in reality, I've hyperlinked that and it's also explained earlier.}

In addition, participants reported that the actual context needed to review documentation comes from confidential internal systems such as Confluence and JIRA, which they cannot share with LLM providers whose retention policies the organization cannot audit or override. \pidf{P11} shared, \qpwi{We have a lot of customer data, like PII data. If there's a data leak, the company also becomes liable... I think something that we always need to tell ourselves every time we use ChatGPT is, are we putting in content that might go out to the world that's not supposed to be for the public eye?} Without access to those sources, LLMs generate feedback from training data such as archived documentation or general internet content, rather than the organization's current product information.

Participants reported that even under the conditions where LLMs lack the necessary context to identify gaps in documentation, they might respond that content is complete and accurate. Participants described this sycophantic tendency as a reason to distrust LLM feedback. For example, \pidf{P2} expressed caution on using LLMs for documentation review, \qpwi{[LLMs] are very people pleasing and will say, `Yes, of course, it looks great. That looks like it has all of the relevant context,' when of course the LLM has no idea what it's missing because it doesn't have that information available to it... An LLM can only act on positive matches in my experience. It doesn't know what to look for if something is missing.}

\subsection{Existing Strategies} 
\label{subsec:existing_strategies}
Writers hold no authority to compel review (\S\ref{subsubsec:process_challenge_prioritizing}), so they must obtain it by negotiating, persuading, or even coercing the reviewers, drawing on their own skill and standing in the organization. In this section, we describe the strategies reported by the participants to \emph{work out}~\cite{corbin_articulation_1993} the arrangements in the documentation review process. While working out is treated as the ordinary business of collaboration in CSCW literature, in documentation review, it substitutes for a mechanism that never existed, since nothing in the process requires reviewers to participate at all. These strategies keep the review process running, but they do not improve it, since they do not leave a trace of the shortcomings that would prompt the organization to fix them~\cite{gasser_integration_1986}.

\subsubsection{Strategic Communication \pc{17}} \textbf{\hlc{(Proactive\slash Corrective, Collaborative)}}
\label{subsubsec:tool_strat_communication}
Participants described three ways of adapting their review requests, based on \emph{when} reviewers were available, \emph{what} they knew, and \emph{how} they preferred to communicate. These suggest that writers reduce the time and attention a review demands of reviewers, since they cannot make them prioritize a review. Participants aimed to increase reviewer engagement by distributing the review requests throughout the code development timeline rather than requesting feedback only in the end. They initiated reviews early in the development process by opening pull requests with specific questions, then coordinated multiple review cycles so reviewers could engage with documentation incrementally as code and corresponding documentation evolved, as described by \pidf{P17}, \qpwi{I'm a believer in opening the PR quite early... I'll send one or two product owners and developers what I'm working on with a few specific questions commented and just be like `Hey, this is what it is right now. I left some questions for you. When you get a chance, can you go through this and answer.' That sets off a back and forth.} This approach ensured reviewers developed familiarity with the content before final approval was needed. \pidfv{P5} noted that early engagement also builds the working relationship that makes substantive review possible; engaging reviewers later in the development cycle becomes progressively harder (\S\ref{subsubsec:process_challenge_prioritizing}, \S\ref{subsubsec:process_challenge_planning}). They also observed that some writers publish partial content and update it as resources allow, rather than waiting until all gaps are addressed. 

Participants also made the review requests highly specific by asking precise, actionable questions that leveraged individual reviewers' technical expertise rather than submitting general requests. This targeted approach reduced cognitive burden on reviewers by giving them clear tasks rather than open-ended input, which increased the likelihood of receiving useful and relevant feedback. For example, \qp{I try and ask specific questions that only an engineer could answer, and that's usually `Where does the information go after this step happens?' or `How does the user know that we're not scamming their credit card data?'}{P17} When participants needed immediate clarification on reviews on complex technical concepts or resolve misunderstandings, they preferred real time interaction through live reviews, screen sharing sessions, and in person meetings over asynchronous communication like GitHub issues or messaging channels. For example, \qp{Getting on a call is the best thing to do because chats create confusion. We have daily standups with the developers, so I join those standups...  We have a discussion, sometimes it takes a day or two... but eventually we agree on something.}{P15}

\subsubsection{Institutional and Social Pressure \pc{12}} \textbf{\hlc{(Proactive\slash Corrective, Collaborative)}}
\label{subsubsec:tool_strat_pressure}
We observed that participants used institutional and social mechanisms to enforce documentation review when their colleagues did not easily provide documentation reviews. Because writers hold no authority of their own to compel review (\S\ref{subsubsec:process_challenge_prioritizing}), these mechanisms work by borrowing authority; participants escalated to upper management to either set boundaries and explicit timelines to ensure documentation received appropriate attention or make participation an expected component of the job. Others tried to persuade colleagues by communicating business impact or emphasizing mutual benefits as in the case of \pidf{P22}, \qpwi{Sometimes the mentality is, `Oh, we'll get it in the next revision.' My response is `OK, but why not just get it now and save our future selves some time? Because I doubt that you are going to remember a year from now to do this portion cause I'm not gonna remember a week from now.' I always try to present what I need in a way that is mutually beneficial.} 

In addition, participants used social pressure by conducting discussions in public channels where both contribution and lack of participation became visible to leadership and peers, which made documentation review into a visible professional responsibility---e.g., \qp{We discuss documentation publicly... in that [Slack] channel with 150 different people [including] PMs, analysts, CTO, and other people who are involved... rather than in DMs. We tag people individually to say your ticket is not ready and if you don't give me feedback, if you don't respond to my questions, it will not be included in release... [Reviewers] see other people getting called out, named and shamed in front of their superiors, but also lots of other people getting praised when a release is good.}{P14} \citet{star_layers_1999} note that work becomes visible only through some indicator of it. Since the review process does not record whether reviewers participated, writers produce that indicator themselves by naming the reviewer in a public channel. 
\section{Discussion}
\label{sec:discussion}
Prior research on coordination suggests that when organizational processes lack a formal mechanism for managing responsibilities, the burden defaults to whoever is most invested in the outcome~\cite{malone_interdisciplinary_1994, crowston_coordination_1997, grinter_supporting_1996}; in documentation review, that burden falls on writers, who are accountable for documentation quality and therefore manage review through their social capital (\S\ref{subsec:existing_strategies}). If that standing erodes or the writer leaves~\cite{robillard_turnover_2021}, the process collapses, and with it the documentation quality. So how do we better support technical writers in this process? In rest of this section, we examine what organizational effort is needed, where tool gaps are most consequential, and what design directions remain open.

\subsection{Practitioners Lack Shared Quality Standards and Incentives to Review}
\label{subsec:doc_review_hard}
Our participants describe that reviewers check documentation against different quality aspects (\S\ref{subsubsec:doc_review_overview}). Therefore, writers must consolidate feedback from all reviewers and judge which inputs to prioritize across quality dimensions. However, writers have no shared standard to guide these judgments. Organizations maintain style guides that serve as a \emph{common information space}~\cite{schmidt_taking_1992} for presentation quality (\S\ref{subsubsec:edit_review}), but nothing comparable exists for content quality or user experience quality. \citet{geiger_types_2018} observed the same pattern in open source data analytics documentation. Without such a standard, writers have no principled basis for deciding whose feedback takes precedence. Participants also report that reviewers often participated minimally or reviewed superficially (\S\ref{subsubsec:process_challenge_prioritizing}). For most reviewers, documentation is not their primary responsibility, so review has no clear connection to their core work. Prior work identifies \emph{task identification}, meaning having goals connected to one's primary work, as the most consistently cited motivator for software practitioners~\cite{beecham_motivation_2008, sharp_models_2009}; reviewers who lack this connection have no intrinsic reason to engage carefully. While writers counter this through institutional escalation and social pressure (\S\ref{subsubsec:tool_strat_pressure}), these strategies mainly compel participation without ensuring quality. When reviewers lack an independent basis for judgment, they tend to align with prior reviewers' feedback rather than assess content on its own merits~\cite{thongtanunam_review_2021}, so the feedback writers arbitrate carries little independent judgment to begin with. Organizations should therefore extend the common information space and thereby establish clear quality criteria for documentation that can give reviewers a concrete basis for independent assessment while providing writers a shared standard for arbitrating conflicting feedback.

\subsection{Organizations Do Not Recognize Writers' Coordination Work}
\label{subsec:invisible_work}
Technical writers do not receive organizational support for coordinating documentation reviews because the work is not tracked which makes it invisible to organizations~\cite{star_layers_1999, deng_investigating_2023, li_crossdisciplinary_2017, meluso_invisible_2025}. The strategies our participants developed in response to challenges in documentation review (\S\ref{subsec:existing_strategies}), such as persuading developers to review or negotiating feedback, are a form of \emph{articulation work}, meaning the effort of putting together tasks, task sequences, and contributions from different workers to keep work flowing~\cite{strauss_articulation_1988}. Articulation work becomes visible only when it breaks down, which is why the challenges in \S\ref{sec:challenges_review} are also the coordination efforts writers otherwise perform silently as \emph{unsung actors} whose contribution is rarely recognized. As such, project tracking tools like JIRA or repository management systems like GitHub do not record this work~\cite{trinkenreich_hidden_2020}. Prior research on coordination mechanisms shows that coordinating distributed work requires a persistent record of each task, its owner, and its current status~\cite{schmidt_coordination_1996}. Because this work is not tracked, organizations cannot measure how much effort writers spend coordinating reviews. This coordination cost further accumulates in the form of \emph{scalar debt} which is the maintenance work whose burden grows with the project but whose infrastructure does not grow to match it~\cite{geiger_labor_2021}. Evidence from open source communities shows that explicit attribution systems for non-code contributions make such work visible~\cite{young_which_2021}. In addition, organizations should also following treat technical writers as \emph{articulation hubs} that organizations formally recognize and support rather than positions that emerge by default~\cite{liu_articulation_2023}, and make efforts to track coordination work alongside documentation related changes.

\subsection{Existing Tools Overlook the Collaboration Aspect of Reviews}
\label{subsec:tooling_support}
\begin{figure}
    \centering
    \includegraphics[width=\linewidth]{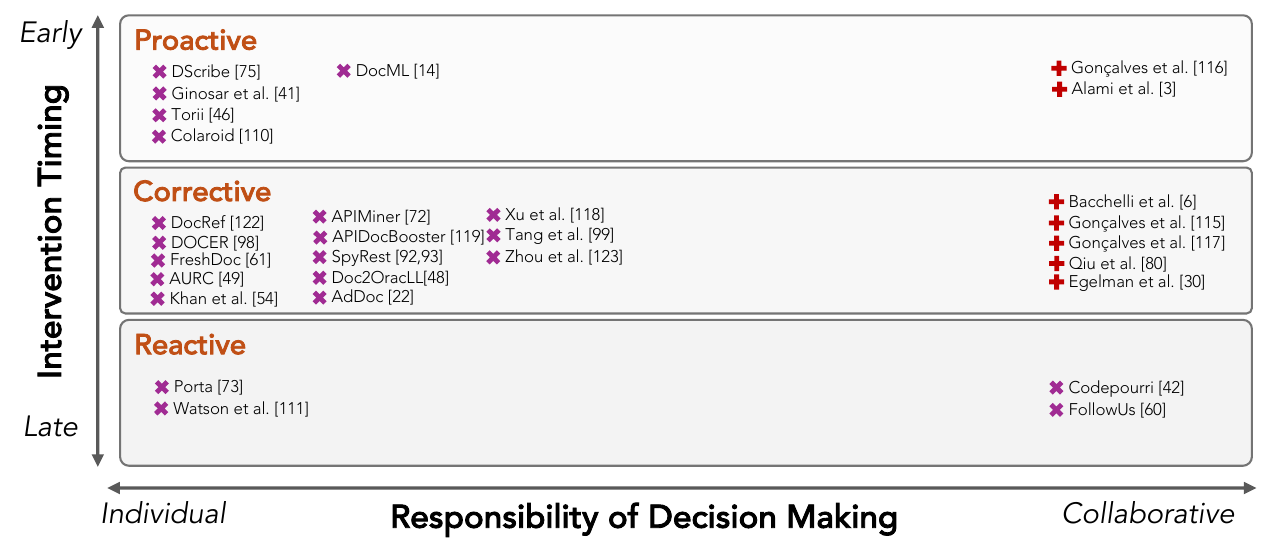}
    \caption{Existing research mapped to Intervention Timing and Responsibility of Decision Making axes. The Collaborative dimension mainly consists of papers focusing on code review (+).} 
    \label{fig:design_space}
\end{figure}

Existing documentation tooling research neglects the collaborative aspect of documentation review, leaving writers to adapt tools built for development workflows (\S\ref{subsubsec:tool_challenge_dedicated}). The docs-as-code framework~\cite{berger_implementing_2024}, defined by the Write the Docs community as ``following the same workflows as development teams, and being integrated in the product team,''~\cite{wtd_docs_as_code_2024} illustrates this. Admittedly, the framework provides benefits like version control and tighter integration with engineering pipelines, but its underlying tools were designed for developers, which excludes non-technical practitioners like editors and product managers from the review process and leaves documentation-specific needs unsupported, such as evaluating rendered HTML output or tracking reviewer consensus (\S\ref{subsubsec:process_challenge_planning}, \S\ref{subsubsec:tool_challenge_maintenance}). We further leverage our findings on characterizing tools used to assist documentation quality as a tool design space~\cite{shaw_role_2012, dove_argument_2016} to identify gaps in existing work on documentation quality and propose viable alternatives. We illustrate it in Fig.~\ref{fig:design_space}.

\noindent \textbf{\hlc{(Corrective, Individual)}.} Most existing tools for documentation quality fall in this category. For content quality, tools like DocRef~\cite{zhong_detecting_2013}, FreshDoc~\cite{lee_automatic_2021}, DOCER~\cite{tan_wait_2023} and others~\cite{hu_aurc_2023, xu_identifying_2025, zhou_analyzing_2017} focus on detecting inconsistencies between code and documentation. Other tools suggest improvements through code examples and summaries (APIMiner~\cite{montandon_documenting_2013}, APIDocBooster~\cite{yang_apidocbooster_2025}, and SpyRest~\cite{sohan_spyrest_2015, sohan_automated_2017}), verify accuracy using test cases and traceable links (Doc2OracLL~\cite{hossain_doc2oracll_2025}, AdDoc~\cite{dagenais_using_2014}), or detect structural and stylistic problems through rule-based systems or trained models~\cite{khan_automatic_2021, tang_evaluating_2023}. Research has likely focused on this category because accuracy is the only quality aspect with a clear ground truth to compare against, namely source code, that aspects like language or style lack.

\noindent \textbf{\hlc{(Proactive, Individual)}.} Fewer tools exist here because proactive quality judgments require context-sensitive human judgment about whether content, language, and structure work for the intended audience. Automated tools cannot encode such judgment as programmable rules~\cite{nassif_generating_2022, ginosar_authoring_2013, head_composing_2020, wang_colaroid_2023, bhat_aspirations_2023}. Admittedly, LLMs can approximate some of these judgments, handling domain nuance and varying their outputs for different users in ways static rule-based systems cannot. However, our findings reveal that current LLM implementations produce unreliable outputs and raise data security concerns in documentation contexts (\S\ref{subsubsec:tool_challenge_reliability}), because writers cannot provide the model with the full documentation or the internal context a review requires. Future work should address these limitations before LLMs can support this area of the design space.

\noindent \textbf{\hlc{(Reactive, *)}.} Existing tools in this category collect explicit user feedback~\cite{watson_tool_2016, dubois_tell_2017, gordon_codepourri_2015, lafreniere_community_2013} and system information as users execute documentation~\cite{mysore_porta_2018}. However, as users increasingly consume documentation through LLMs~\cite{raglianti_on_2023}, the analytics and comments that official websites once enabled become obsolete, and organizations can no longer see how documentation is used (\S\ref{subsubsec:reactive_timing}). These analytics were the \emph{indicators}~\cite{star_layers_1999} through which organizations saw documentation quality (\S\ref{subsec:doc_review_hard}). However, as documentation is increasingly mediated through LLMs, these indicators disappear; organizations lose the feedback of how well documentation serves users, which can result in eventual decline of documentation quality~\cite{bhat_who_2026}.

\noindent \textbf{\hlc{(*, Collaborative)}.} The collaborative dimension is the most significant gap in the design space, and the least studied. Code review, by contrast, has evolved into a rich area of research~\cite{eldh_code_2024, badampudi_modern_2023}, and researchers have developed tools to support collaboration directly, including recommending reviewers, visualizing changes, annotating code, and automating feedback~\cite{davila_systematic_2021, tufano_towards_2021, tufano_code_2024, sarkar_automated_2023}. This body of work draws on empirical research on what collaboration in review processes involves, including how practitioners negotiate feedback, manage conflicting priorities, and navigate interpersonal dynamics~\cite{bacchelli_expectations_2013, wurzel_competencies_2023, alami_accountability_2025, goncalves_interpersonal_2022, egelman_predicting_2020}. Documentation review involves the same dynamics (\S\ref{subsubsec:process_challenge_prioritizing}, \S\ref{subsubsec:tech_review}), yet without equivalent empirical work, tool designers have had no basis for addressing them. Additionally, a collaborative review tool does not by itself guarantee coordination. Its users still need a shared standard to review against and an obligation to take part~\cite{schmidt_taking_1992}, which documentation review lacks (\S\ref{subsec:doc_review_hard}, \S\ref{subsubsec:process_challenge_prioritizing}). By showing what that coordination consists of and where it breaks down, our findings provide the empirical basis that this dimension has lacked, making it actionable for tool designers.

\subsection{Designing Intelligent Tools to Support Collaboration in Reviews}
\label{subsec:ai_implications}
We suggest two design directions to support writers in the \textbf{\hlc{(*, Collaborative)}} dimension of the design space; reducing the cognitive burden that produces shallow reviewer engagement, and facilitating collaboration with non-technical reviewers. Our participants report that reviewers disengage from documentation review because it falls outside their primary responsibilities, producing superficial approvals that leave writers unable to tell whether feedback was substantive (\S\ref{subsubsec:process_challenge_prioritizing}, \S\ref{subsubsec:process_challenge_expertise}). An intelligent assistant could reduce this burden by providing each reviewer with context matched to their role and generating targeted, role-specific questions before they engage~\cite{woolley_generative_2025}, shifting the work of framing the review task off the writer. 

To include non-technical reviewers, we suggest a collaborative environment where each practitioner works in a space matched to their reviewing task without requiring them to learn developer tooling (\S\ref{subsubsec:tool_challenge_dedicated}). Non-technical reviewers are excluded when documents are shown in raw markdown as on GitHub, so they would be better served by the rendered form (\S\ref{subsubsec:tool_challenge_reliability}). Malleable software makes this feasible by letting practitioners reshape interfaces to fit their work rather than adapting their work to fixed tools~\cite{tchernavskij_designing_2019, klokmose_webstrates_2015, klokmose_mywebstrates_2024}. Future work can leverage design principles for how such systems should be structured, supporting interaction where each user's view is organized around its own objects and constraints~\cite{mackay_interaction_2025, beaudouin-lafon_generative_2021}. While these directions make the writer's coordination work visible (\S\ref{subsec:invisible_work}), we add one cautionary note, that these technologies must be designed to empower writers since record of coordination can otherwise be used to monitor writers and eventually pressure them to perform the work that only improves the metrics rather than being useful~\cite{star_layers_1999, suchman_making_1995}.

\section{Conclusion}
In this work, we sought to bridge the gap between knowing what constitutes high-quality documentation and understanding how it is achieved. Our investigation reveals that documentation quality emerges through a coordinated review process involving multiple practitioners with complementary expertise. However, this process remains inadequately supported by existing workflows and tools, in part because the expertise and coordination it requires are invisible to the systems organizations use to evaluate work. Technical writers, who orchestrate these review processes, must develop workarounds to manage related challenges. Informed by our findings, we propose a design space to help researchers and professionals diagnose any process failures and systematically explore alternative mechanisms to support documentation review. We call for future research to investigate purpose built collaborative tools and explore how emerging technologies like LLMs can augment human expertise and collaboration in these workflows.
\bibliography{bibliography/references.bib}
\bibliographystyle{ACM-Reference-Format}
\end{document}